\documentclass[twocolumn,twocolappendix]{aastex631}
\usepackage{lineno}
\usepackage{longtable}

\newcommand{\atob}{\text{--}}
\newcommand{\keV}{\mathrm{keV}}
\newcommand{\Chandra}{\textit{Chandra}\ }
\newcommand{\XMM}{\textit{XMM-Newton}\ }
\newcommand{\MeerKAT}{\textit{MeerKAT}\ }
\newcommand{\WISE}{\textit{WISE}\ }
\newcommand{\Gaia}{\textit{Gaia}\ }
\newcommand{\kpc}{\mathrm{kpc}}

\newcommand{\logLx}{\log (L_\mathrm{X}/ \mathrm{erg\ s^{-1}})}

\newcommand{\Msun}{\mathrm{M_\odot}}
\newcommand{\s}{\mathrm{s}}

\newcommand{\ergs}{\mathrm{erg\ s^{-1}}}
\newcommand{\phcms}{\mathrm{ph\ cm^{-2}\ s^{-1}}}

\newcommand{\Sband}{$S$-band}
\newcommand{\Hband}{$H$-band}
\newcommand{\Fband}{$F$-band}

\newcommand{\barRe}{\bar{Re}}

\shorttitle{An X-Ray Source Catalog on Antlia}
\shortauthors{Zhensong Hu}

\graphicspath{{}{}}

\usepackage{enumitem} 
\begin{document}

\title{AMUSE-Antlia. II.  Intracluster X-ray Population in the Antlia Cluster}

\author[0009-0009-9972-0756]{Zhensong Hu}
\affiliation{School of Astronomy and Space Science, Nanjing University, Nanjing 210023, China}
\affiliation{Key Laboratory of Modern Astronomy and Astrophysics (Nanjing University), Ministry of Education, Nanjing 210023, China}
\email{huzhensong@smail.nju.edu.cn}

\author[0000-0002-3886-1258]{Yuanyuan Su}
\affiliation{Department of Physics and Astronomy, University of Kentucky, 505 Rose street, Lexington, KY 40506, USA}
\email{ysu262@g.uky.edu}

\author[0000-0003-0355-6437]{Zhiyuan Li}
\affiliation{School of Astronomy and Space Science, Nanjing University, Nanjing 210023, China}
\affiliation{Key Laboratory of Modern Astronomy and Astrophysics (Nanjing University), Ministry of Education, Nanjing 210023, China}
\affiliation{Institute of Science and Technology for Deep Space Exploration, Nanjing University, Suzhou 215163, China}
\email{lizy@nju.edu.cn}

\author[0000-0001-9062-8309]{Meicun Hou}
\affiliation{Institute of Science and Technology for Deep Space Exploration, Nanjing University, Suzhou 215163, China}

\author[0000-0002-0765-0511]{Ralph P. Kraft}
\affiliation{Harvard-Smithsonian Center for Astrophysics, 60 Garden Street, Cambridge, MA 02138, USA}

\author[0000-0001-9662-9089]{Kelley M. Hess}
\affiliation{Department of Space, Earth and Environment, Chalmers University of Technology, Onsala Space Observatory, 43992 Onsala, Sweden}
\affiliation{ASTRON, the Netherlands Institute for Radio Astronomy, Postbus 2, 7990 AA, Dwingeloo, The Netherlands}

\author[0000-0002-3612-9258]{Hao Chen}
\affiliation{Research Center for Intelligent Computing Platforms, Zhejiang Laboratory, Hangzhou 311100, China}

\begin{abstract}

We conduct a systematic survey of  X-ray sources in the inner ($r\sim200$ kpc) region of the Antlia cluster based on \Chandra observations, down to a source detection limit of $ L(0.5\text{--}8\ \keV)\sim4.2\times10^{-7}\ \phcms$ ($2\times10^{38}\ \ergs$).  We present an X-ray source catalog with 202 sources and provide their coordinates, multi-band flux information and hardness ratios. We find a statistically significant excess at a significance level of $4.2\sigma$ with 37.6 excess sources beyond three times the mean effective radius of the two BCGs. This implies that these excess sources could be a genuine intracluster X-ray population that is not associated with the bulk stellar component. Also, the increased number of excess sources in the fields containing a BCG implies a potential connection between the excess sources and BCGs. The discovery of these sources in the Antlia cluster, together with previous research of similar findings in other two nearby clusters, Virgo and Fornax, indicates that the intracluster X-ray population could be universal in nearby galaxy clusters. Furthermore, 
we discuss the candidate origins of the excess sources, including low-mass X-ray binaries (LMXBs) associated with intracluster light (ICL-LMXBs),  LMXBs in globular clusters (GC-LMXBs) and supernova-kicked LMXBs (SN-kicked LMXBs). We estimate the contribution of ICL-LMXBs, which should include the LMXBs relating with the stellar halo surrounding BCGs, are unlikely to dominate  the intracluster X-ray population in Antlia. Meanwhile, GC-LMXBs and SN-kicked LMXBs, each component could contribute $\sim30\%$ to the total excess sources. 

\end{abstract}


\section{Introduction} \label{sec:intro}

X-ray binaries (XRBs) in nearby galaxies are thought to be a good tracer of the bulk stellar component \citep{2006ARA&A..44..323F}.  High-mass X-ray binaries (HMXBs), which accrete stellar wind from its high-mass optical companion, are closely associated with the star formation rate of the host galaxy \citep{2003MNRAS.339..793G}. For low-mass X-ray binaries (LMXBs) that accrete material through the Roche-lobe flow, they are tightly related with the old stellar population. In particular, the total number of LMXBs and their combined luminosity are proportional to the host stellar mass \citep[e.g.][]{2004MNRAS.349..146G,2004ApJ...611..846K}.  
However, this relation is biased in the outskirts of galaxies. X-ray sources, down to $L_\mathrm{X}\gtrsim10^{37}\ \ergs$, have been detected in the  halo of several massive early-type galaxies (ETGs), where the stellar component is negligible, such as the Sombrero Galaxy (M104) \citep{2010ApJ...721.1368L} as well as a broad sample of 20 ETGs \citep{2013A&A...556A...9Z}. Both studies suggest two origins could be equally important for the excess of halo X-ray sources: LMXBs harbored in globular clusters (GC-LMXBs) and those kicked out during the anisotropic supernova explosion of a compact object (SN-kicked LMXBs).

Recent studies on two nearby clusters, Virgo and Fornax, reveal a new group of X-ray sources that relates with the galaxy cluster environment. 
For the Virgo cluster, \citet{2017ApJ...846..126H} have combined the X-ray source profiles (down to a luminosity limit of $L_\mathrm{X}\gtrsim 2\times10^{38}\ \ergs$) of 80 Virgo low-to-intermediate-mass ETGs, and have found an excess with a significance of $3.5\sigma$ and $\sim120$ excess sources in the halos of (outside three times effective radii) member galaxies. On the other hand, such an excess has not been found in their field analogs. This led \citet{2017ApJ...846..126H} to propose the presence of the so-called intracluster X-ray population (ICX), which is defined as X-ray sources that are gravitationally unbound to any member galaxies. There could be multiple origins of ICX. For example, the diffuse intracluster light (ICL) which is conventionally detected in the optical/near-infrared band \citep{2022NatAs...6..308M}, 
that is dominated by old stellar population, as shown by its red color \citep[e.g.][]{2017ApJ...834...16M, 2022NatAs...6..308M},  can naturally harbor LMXBs (hereafter ICL-LMXBs).  Also, GC-LMXBs and SN-kicked LMXBs may  have a substantial contribution. 
Subsequently, in a recent study focusing on Virgo late-type galaxies (LTGs), \citet{2024ApJ...968...41H} have found 20 excess X-ray sources at $\sim3.3\sigma$ significance in the combined X-ray sources surface density profiles of 19 edge-on Virgo LTGs. They have suggested that newly-formed HMXBs, associated with the enhanced star formation rate caused by ram pressure, as a source of ICX around cluster LTGs.
In addition, another survey targeted at the Fornax cluster have revealed 183 ICX sources with a maximum significance of $3.6\sigma$ beyond three times effective radius of the brightest central galaxy (BCG) NGC\,1399, down to a luminosity sensitivity of $L_\mathrm{X}\gtrsim 3\times10^{37}\ \ergs $  \citep{2019ApJ...876...53J} . 
Yet, current researches are limited in only two clusters, therefore, an extension of samples is needed to confirm whether the ICX is universal in nearby clusters.

Antlia is the third nearest cluster (35.2\ Mpc, $1''=170\ $pc) \citep{2003A&A...408..929D} after Virgo and Fornax. The main group, centered at the BCG NGC\,3268, is merging with another subgroup centered at NGC\,3258. Antlia has a virial radius of $887\ $kpc and a virial mass of $7.9\times10^{13}\ \Msun$\citep{2016ApJ...829...49W}, which are comparable with those of Virgo and Fornax.  In this work,  we present a systematic analysis on the X-ray sources and look for ICX in Antlia. This paper is structured as follows. Section \ref{sec:method} describes the procedures for data preparation and X-ray source detection, then an X-ray source catalog is displayed in Section \ref{sec:Catalog}. In Section \ref{sec:ICX} we search for Antlia ICX and Section \ref{sec:discussion} discusses its potential origins. Lastly, Section \ref{sec:summary} summarizes this paper.

\section{Data Processing}\label{sec:method}

\subsection{Data Preparation}\label{subsec:data_preparation}

We observed the central region ($\sim 200\ \kpc$) of the Antlia cluster during June 2021 to March 2022, with three Chandra ACIS-I pointings (PI: Y. Su). These three fields were targeted at the BCG NGC\,3268, the southwest group centered on another BCG, NGC\,3258, and the southeast of the cluster. Each field has an approximately $ 70 \ \mathrm{ks} $ exposure, and the total exposure time is $ 223.89\ \mathrm{ks} $.  The \Chandra observation IDs, pointing coordinates, exposures, and other information are listed in Table \ref{table:obsinfo}.

\begin{deluxetable}{cccccc}
\tablenum{1}
\tablecaption{Log of Chandra Observations \label{table:obsinfo}}
\tablewidth{0pt}
\tablehead{
\colhead{ObsID}  &\colhead{Exposure} & \colhead{R.A.} &
\colhead{Decl.} & \colhead{StartDate} & Field\\ & (ks) & (deg) & (deg) & &}
\decimalcolnumbers
\startdata
23794                                                                                           & 55.35                                                  & 157.529125                                       & -35.290086                                   & 2022 Jan 21          & 3268                                          \\
24731                                                                                           & 19.64                                                  & 157.529125                                       & -35.290086                                      & 2022 Jan 3    & 3268                                           \\
23795                                                                                           & 26.46                                                  & 157.182417                                     & -35.582019                                      & 2022 Mar 18    & 3258                                          \\
24732                                                                                           & 21.1                                                   & 157.182417                                     & -35.582019                                      & 2021 Jun 20    & 3258                                             \\
26361                                                                                           & 25.74                                                  & 157.182417                                     & -35.582019                                      & 2022 Mar 20     & 3258      \\
23796                                                                                           & 17.74                                                  & 157.564167                                     & -35.592958                                     & 2021 Oct 14    & SE                                          \\
24733                                                                                           & 19.81                                                  & 157.564167                                     & -35.592958                                     & 2021 Jul 15        & SE                                      \\
26157                                                                                           & 17.35                                                  & 157.564167                                     & -35.592958                                     & 2021 Oct 15      & SE                                        \\
26158                                                                                           & 12.73                                                  & 157.564167                                     & -35.592958                                     & 2021 Oct 15    & SE                                           \\
26160                                                                                           & 7.97                                                   & 157.564167                                     & -35.592958                                     & 2021 Dec 9     & SE               
\enddata
\tablecomments{($ 1 $) Observation ID. All observations are based on \Chandra ACIS-I detectors. ($ 2 $)  Exposure time, in units of $ \mathrm{ks} $. ($ 3 $)$ - $($ 4 $) Epoch 2000 coordinate of the aimpoint. ($ 5 $) Start date of the observation, in the form of yyyy-mm-dd. ($6$) The targeted field, where ``3268", ``3258" and ``SE" stand for the fields of NGC~3268, NGC~3258 and  southeast, respectively. }
\end{deluxetable}

Following the official procedures of CIAO v$4.14$ \citep{2006SPIE.6270E..1VF}, we reprocessed the level-$1$ data  and the attached calibration files. Then we corrected the astrometry for all observations with the CIAO tool \texttt{reproject\_aspect}, by setting the longest-exposed image of each field as the reference image.  After that, we inspected the lightcurve of each observation and found that the particle background is quiescent, so we kept all the scientific data for further analysis. 

\begin{figure*}[ht!]
\epsscale{1.2}
\plotone{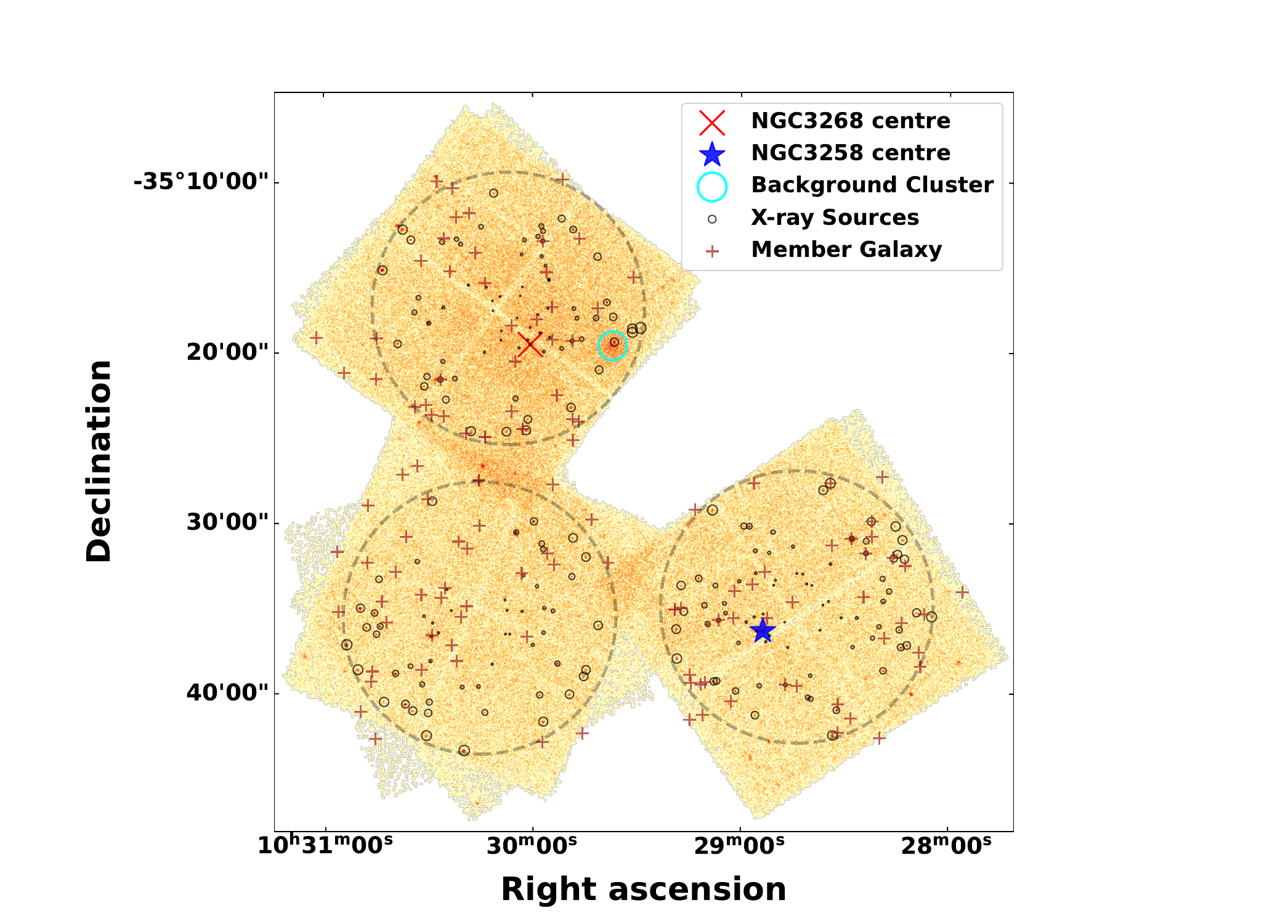}
\caption{The mosaic $0.5-8\ \mathrm{keV}$ counts image of the Antlia cluster, smoothed by a $ 2 $ pixel Gaussian kernel. The centers of NGC $ 3268 $ and NGC $ 3258 $ are marked by a red ``$\times$"  and a blue ``$\star$", respectively. Other member galaxies are marked by a brown ``$+$". The small black circles indicate the X-ray point sources, and their radii are twice the 90\% ECR. 
The background cluster at redshift $z=0.42\pm0.01$ \citep[e.g.][]{2018MNRAS.479..240G} westward of the NGC\,3268 field is labelled with a cyan circle.
The gray dashed circles with a fixed radius of $8'$ are centered at the aimpoint of each FoV.  \label{fig:X-ray_map}}    
\end{figure*}

We created counts maps and exposure maps at a pixel scale of $0.''492$ in three energy bands: $0.5-2 \ \keV$ ($S$-band), $2-8\ \keV$ ($H$-band), and $0.5-8 \ \keV$ ($F$-band). The exposure maps were weighted by a fiducial incident spectrum of an absorbed power-law, with a photon index of $1.7$ and an absorption column density of $N_\mathrm{H} = 10^{21}\ \mathrm{cm^{-2}}$. The latter value is somewhat higher than the actual Galactic foreground absorption column density ($\sim 6.9\times10^{20} \ \mathrm{cm^{-2}}$), but accounts for some internal absorptions in LMXBs \citep[e.g.][]{2018ApJ...862...73L}. The point-spread function (PSF) map was generated at a given enclosed count radius (ECR) individually for each observation ID in the three energy bands. At last, we created a merged event list and a merged exposure map for each field by combining the event lists and exposure maps of the observations having the same pointings. The merged PSF maps were weighted based on the local effective exposure. Figure \ref{fig:X-ray_map} is a schematic view of the total counts image in the $F$-band, smoothed by a 2-pixel Gaussian kernel.  

\subsection{Source Detection}\label{subsec:src_detection}

We followed the source characterization method in \citet{2018ApJS..235...26Z} and \citet{2019ApJ...876...53J}. The same process was applied individually on the combined products of each field. The key steps and procedures are summarized here. 

Firstly, we utilized the CIAO tool {\tt wavdetect} to perform source detection in the three energy bands,  with a false-detection probability threshold of $ 10^{-6} $, using the combined exposure maps and the $ 50\% $ ECR PSF maps.
Secondly, we refined the source centroids by iterating over the source positions within the $90\%$ ECR. Thirdly, we calculated the position uncertainties of the X-ray sources (PU$_{\rm X}$) at a $ 68\% $ confidence level according to an empirical relation between PU$_{\rm X}$, source counts and source off-axis angle \citep{2007ApJ...659...29K}.  Then all sources that are beyond $ 8' $ from the aimpoint  were removed,   ensuring that all sources are covered by multiple observations with the same aimpoints but different roll angles. 
The photon fluxes were measured by extracting source counts within the 90\% ECR, and the background was calculated based on an annulus with inner-to-outer radii of 2–4 times the 90\% ECR. For less luminous sources, we adopted a photon-to-energy conversion factor of $3.29\times10^{-9}\ \mathrm{erg\ ph^{-1}}$ according to the assumed fiducial incident power-law.  On the other hand, for bright sources with more than 100 net counts ($S_{0.5-8\ \keV} \gtrsim 1.3\times10^{-5}\  \phcms$), we derived  energy fluxes by spectrum fitting, assuming an absorbed power-law spectrum with a fixed absorption column density of $N_\mathrm{H} = 10^{21}\ \mathrm{cm^{-2}}$ and a free photon index.  
After that, to filter out spurious sources caused by background fluctuation, we calculated the binomial no-source probabilities $P_\mathrm{B}$  (\citealt[Appendix A2]{2007ApJ...657.1026W}; \citealt[Equation 1]{2018ApJS..235...26Z}). Any sources with $P_\mathrm{B} > 0.01$ were considered spurious and excluded.

Finally, we employed a cross-matching method \citep{2009ApJ...706..223H} to identify the same source detected in different energy bands. We defined the relative distance $ d_r = d_{12}/\sqrt{\mathrm{PU}_1^2+\mathrm{PU}_2^2}$, where $d_{12}$ is the absolute distance between the two sources,  and $\mathrm{PU}_1$ and $\mathrm{PU}_2$ are the positional uncertainties of source 1 and source 2. Any two sources with $d_r < 3.0$ were thought to be one source detected in different energy bands. In addition, we tested their random match probabilities by shifting the full band source positions with $10''$ in four directions. As a result, we estimated $0.25$($0.25$) random matches out of the $109$ ($111$) $S/F$ ($H/F$) pairs, i.e., $\sim 0.23\%$ false matches. The rate of false matches is negligible.

Diffuse emission or gas clumps can lead to false detection by \texttt{wavdetect} \citep[e.g.][]{2018ApJ...862...73L}, due to its basic algorithm that looks for ``local peaks''.  So, we additionally checked all sources by eye to identify any false detection.  We found only two sources, both locating at the very center of NGC~3258, that overlap with local strong diffuse emission. A further discussion of these sources can be found in Section 3 of \citet{2023ApJ...956..104H}.

\begin{figure}[ht!]
\epsscale{1.2}
\plotone{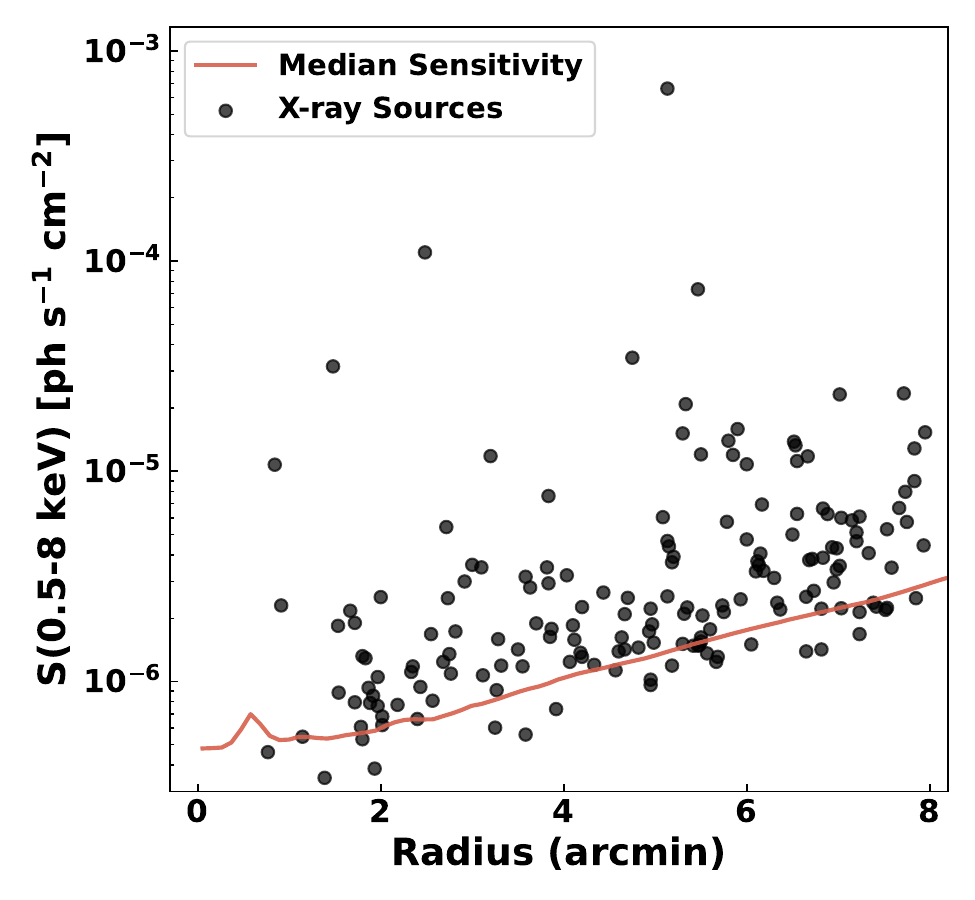}
\caption{The red line indicates the  $ 0.5-8\ \mathrm{keV}$ median sensitivity vs. projected radius from the aimpoint of corresponding field. The black circles represents the $F-$band sources. The sources detected at larger radius are more luminous than inner sources due to sensitivity incompleteness. \label{fig:sensitivity}}
\end{figure}

The source detection sensitivity changes throughout the FoV, due to the different off-axis angles, the background emission, the non-uniform exposure of combined observations and the PSF variation. We produced a sensitivity map for each field with the chosen false-positive threshold ($10^{-6}$, as used for source detection), following the recipe in \citet{2010ApJ...719..900K}. Thus, the flux limit and incompleteness of the Chandra source sample are
derived according to the sensitivity map. In Figure \ref{fig:sensitivity}, we show the azimuthally median sensitivity as a function of the radius $R$ towards the aimpoint, which is derived from the sensitivity map. The small peak at $\sim 0.8'$ is caused by the narrow gaps between CCDs.  
The flux limit of the NGC\,3268 field, the NGC\,3258 field and the  southeast field  are $4.9\times10^{-7} \ \phcms$ ($2.4\times10^{38}\ \ergs$), $4.3\times 10^{-7} \ \phcms$ ($2.1\times10^{38}\ \ergs$), and $4.2\times10^{-7}  \ \phcms$ ($2.0\times10^{38}\ \ergs$), respectively.

\section{Main Catalog}\label{sec:Catalog}

The final source catalog of X-ray sources in the central $\sim200\ \kpc$ region of the Antlia cluster is presented in Table \ref{table:source_catalog}. We identified 202 sources in total, among which $ 125 $ are in the \Sband, $ 123 $ are in the \Hband, and $ 173 $ are in the \Fband.   
In this section, we describe the process of identifying  X-ray source counterparts and analyze the hardness ratios.

\begin{deluxetable*}{cccccccccc}
\tablenum{2}
\tablecaption{Catalog of Detected X-Ray Sources in the Antlia Cluster\label{table:source_catalog}}
\tablewidth{0pt}
\tablehead{\colhead{No.} & \colhead{R.A.} & \colhead{Decl.} & \colhead{PU$_{\rm X}$} & \colhead{ $ S_\mathrm{0.5-2} $ } & \colhead{ $ S_\mathrm{2-8}  $ } & \colhead{ $ S_\mathrm{0.5-8} $ }& \colhead{ $ F_\mathrm{0.5-8} $ } & \colhead{ HR }  & \colhead{Note} }
\decimalcolnumbers
\startdata
$ 1 $ & $ 157.02025 $ & $ -35.59128 $ & $ 0.81 $ & $ 37.0^{+9.1}_{-8.1} $ & $ <25.1 $ &  $ 44.4^{+8.8}_{-8.2} $ & $ 146.1^{+29.0}_{-27.0} $ & $ -0.22^{+0.24}_{-0.19} $ & $\dots$ \\ 
$ 2 $ & $ 157.03836 $ & $ -35.58737 $ & $ 0.87 $ & $ <42.8 $ & $ <13.4 $ &  $ 22.3^{+6.4}_{-5.8} $ & $ 73.4^{+21.1}_{-19.1} $ & $ -0.51^{+0.15}_{-0.49} $ & $\dots$ \\ 
$ 3 $ & $ 157.05003 $ & $ -35.61934 $ & $ 0.47 $ & $ 28.7^{+7.4}_{-6.6} $ & $ 29.3^{+5.0}_{-4.7} $ &  $ 66.5^{+8.9}_{-8.9} $ & $ 218.8^{+29.3}_{-29.3} $ & $ 0.38^{+0.13}_{-0.12} $ & R \\ 
$ 4 $ & $ 157.05306 $ & $ -35.53511 $ & $ 1.13 $ & $ 17.4^{+6.8}_{-5.9} $ & $ <6.3 $ &  $ <28.0 $ & $ <92.1 $ & $ -0.85^{+0.02}_{-0.15} $ & $\dots$ \\ 
$ 5 $ & $ 157.05561 $ & $ -35.51630 $ & $ 0.69 $ & $ 30.7^{+8.4}_{-7.5} $ & $ 13.1^{+4.0}_{-3.6} $ &  $ 40.8^{+8.2}_{-7.6} $ & $ 134.2^{+27.0}_{-25.0} $ & $ 0.10^{+0.18}_{-0.18} $ & $\dots$ \\ 
$ 6 $ & $ 157.05719 $ & $ -35.62126 $ & $ 0.45 $ & $ 36.1^{+8.0}_{-7.3} $ & $ 24.4^{+4.7}_{-4.4} $ &  $ 62.7^{+8.7}_{-8.6} $ & $ 206.3^{+28.6}_{-28.3} $ & $ 0.16^{+0.14}_{-0.13} $ & $\dots$ \\ 
$ 7 $ & $ 157.05917 $ & $ -35.60416 $ & $ 0.48 $ & $ <26.5 $ & $ 21.8^{+4.1}_{-3.9} $ &  $ 40.6^{+7.0}_{-6.6} $ & $ 133.6^{+23.0}_{-21.7} $ & $ 0.66^{+0.15}_{-0.11} $ & $\dots$ \\ 
$ 8 $ & $ 157.06136 $ & $ -35.53022 $ & $ 0.74 $ & $ <35.4 $ & $ 17.4^{+5.0}_{-4.4} $ &  $ 37.8^{+9.2}_{-8.4} $ & $ 124.4^{+30.3}_{-27.6} $ & $ 0.64^{+0.24}_{-0.18} $ & R \\ 
$ 9 $ & $ 157.06390 $ & $ -35.50303 $ & $ 1.57 $ & $ 12.9^{+6.7}_{-5.7} $ & $ <11.4 $ &  $ <31.7 $ & $ <104.3 $ & $ -0.47^{+0.17}_{-0.53} $ & $\dots$ \\ 
$ 10 $ & $ 157.06600 $ & $ -35.53436 $ & $ 0.76 $ & $ <37.0 $ & $ 10.3^{+3.8}_{-3.5} $ &  $ 22.0^{+6.9}_{-6.3} $ & $ 72.4^{+22.7}_{-20.7} $ & $ 0.26^{+0.27}_{-0.23} $ & $\dots$ 
\enddata
\tablecomments{($ 1 $) Source ID, which is listed in ascending order based on the right ascension. Bright sources, with more than 100 net counts ($S_\mathrm{0.5-8} \gtrsim1.3\times10^{-5}\ \phcms$) , was indicated with a marker ``$*$". ($ 2 $)$- $($ 3 $)  Right ascension and declination (J2000) of the source centroid. ($ 4 $) Position uncertainty, in arcseconds. ($ 5 $)$ - $($ 7 $) The $0.5\atob2\ \mathrm{keV}$, $2\atob8\ \mathrm{keV}$ and $0.5\atob8\ \mathrm{keV}$ photon flux, in units of $10^{-7}\ \mathrm{ph\ cm^{-2}\ s^{-1}}$. ($ 8 $) The $0.5-8\ \mathrm{keV}$ unabsorbed energy flux by assuming an absorbed power-law spectrum with a photon index of $1.7$ and column density $10^{21}\ \mathrm{cm^{-2}}$, in units of $10^{16} \ \mathrm{erg \ cm^{-2} \  s^{-1}}$. The fluxes of bright sources, denoted with a symbol ``$*$", are derived from spectral fitting with Xspec based on an absorbed power-law model \texttt{phabs*powerlaw}, assuming a fixed absorption of $10^{21}\ \mathrm{cm^{-2}}$, but the photon index is set to be free.  ($ 9 $) Hardness ratio, defined as $HR=(H-S)/(H+S)$. ($ 10 $) Identification of the sources, where ``R'' means having a radio counterpart, ``F'' denotes a foreground star, and ``N'' denotes nuclear source of a member galaxy.  Quoted errors are at the $ 1\sigma $ confidence level, while $3\sigma$ upper limits are given in the case of non-detection in a given band.  (This table is available in its entirety in machine-readable form.)}
\end{deluxetable*}

\subsection{Counterparts}

\citet{2023ApJ...956..104H} present a comprehensive analysis of AGNs identified through \Chandra X-ray observations in  Antlia, along with a comparison of their nuclear activity with that of the Virgo and Fornax clusters.  Nine galaxies are found to harbor an AGN, namely FS\,82, FS\,88, FS\,98, FS\,105 (NGC\,3257), FS\,125 (NGC\,3260), FS\,168 (NGC\,3267), FS\,184 (NGC\,3269), FS\,185 (NGC\,3268) and FS\,226 (NGC\,3273). The names of the Antlia member galaxies are adopted from the galaxy catalogs from \citet{1990AJ....100....1F}. 

We identify foreground stars by cross-matching our source catalog with \Gaia Data Release 3 \citep{2016A&A...595A...1G,2023A&A...674A...1G}. A foreground star is taken to have a parallax-over error $ > 5 $ because the typical parallax error $0.2\ \mathrm{mas}$ at $G=18\ \mathrm{mag}$ corresponds to a distance of $\sim5\ \kpc$  \citep{2023A&A...674A...1G}. As a result, we find 3 X-ray sources having a \Gaia counterpart, which are labelled with an ``F'' in Table \ref{table:source_catalog}. Also, we obtain a random match test by shifting the source positions by $10''$, resulting in one random matched pair. In a word, one out of the three identified foreground stars may be attributed to a false match.

In addition, we search for any X-ray sources paired with a \MeerKAT radio source.  \MeerKAT observed (SCI-20210212-KH-01; PI: K. Hess) the Antlia cluster in the L band, covering 856–1712 MHz frequency range. The radio continuum image obtains an rms noise of $\sim6.5\ \mu\mathrm{Jy/beam}$ and an angular resolution of $7''$, with an astrometric uncertainty of $1.''5$. The matching radius is chosen as three times the total position uncertainty $PU_{T} = \sqrt{\mathrm{PU_{X}}^2 + \mathrm{PU_{R}}^2}$, where $\mathrm{PU_{X}}$ and $\mathrm{PU_{R}}$ represent the position uncertainty of an X-ray sources and a radio source. Consequently, we find 68 X-ray sources paired with a radio source. Among these sources, 8 are identified AGNs as mentioned in \citet{2023ApJ...956..104H}, and they have a confirmed optical counterpart that corresponds to a known  member galaxy. The random match test result in 1.75 random matched pair. Then we try to look for any of these sources that relates with a galaxy. Even though we find 5 sources located within three times the effective radius of a galaxy (excluding the radio sources paired with an AGN), the random match test which shifts the sources by $100''$ in four directions finds 4.5 random match pairs. Thus, it is unlikely for any of these X-ray sources with a radio counterpart to be related with a certain Antlia member galaxy. These sources may be background AGNs or microquasars. However, the nature of these sources is still unclear and they are still kept in further analysis.

\subsection{Hardness Ratio}

\begin{figure}[ht!]
\epsscale{1.2}
\plotone{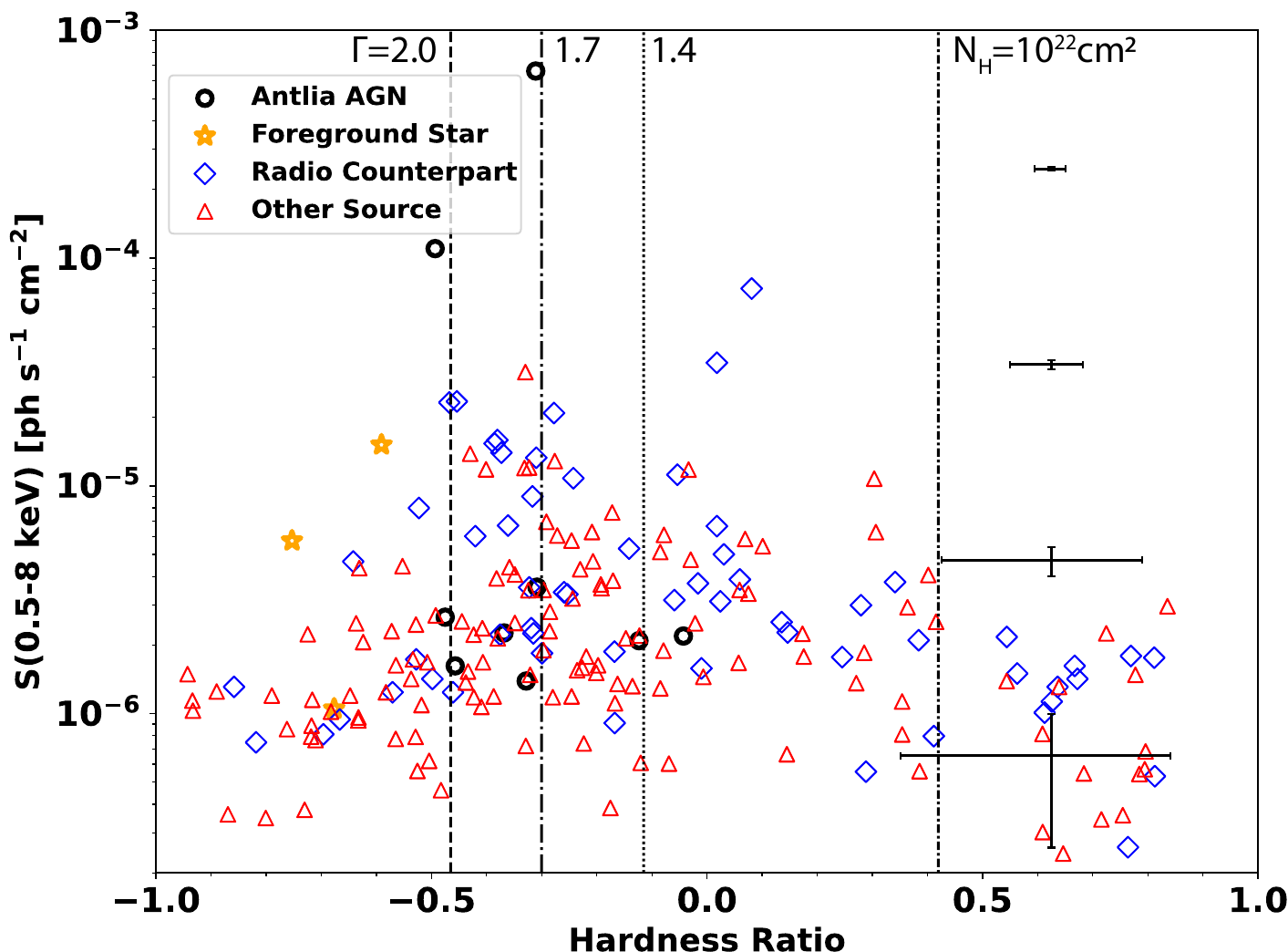}
\caption{$0.5-8\ \mathrm{keV}$ flux vs. hardness ratio.  The black circles, yellow stars, blue diamonds and red triangles represent identified AGNs, foreground stars, X-ray sources with a radio counterpart and other X-ray sources, respectively.  For those eight AGNs with a radio counterpart, they are only denoted as a black circle. The error bars, from top to bottom, denote the median errors of sources with a flux of $S_{0.5-8}\geqslant10^{-4}\ \phcms$, $10^{-5}\leqslant S_{0.5-8}< 10^{-4}\ \phcms$, $10^{-6}\leqslant S_{0.5-8}<10^{-5}\ \phcms$, and $S_{0.5-8}<10^{-6}\ \phcms$.  
The vertical dashed lines, from left to right, correspond to hardness ratios for an absorbed power-law spectrum with a fixed absorption of $N_\mathrm{H} = 10^{21}\ \mathrm{cm^{-2}}$ and a photon index of 2.0, 1.7 and 1.4. The dashed dotted line on the right indicates the hardness ratio based on an absorbed power-law spectrum with a photon index of 1.7 and an absorption column density of $N_\mathrm{H} = 10^{22}\ \mathrm{cm^{-2}}$. \label{fig:hardnessratio}}
\end{figure}

We define  hardness ratio (HR) as $(H-S)/(H+S)$, where $H$ and $S$ are the net counts in the \Hband\ and the \Sband. Figure \ref{fig:hardnessratio} shows  HR versus  $F$-band  photon flux. The foreground stars have a much softer HR, while other sources, 
generally obtain an HR value consistent with being an AGN or an LMXB, i.e. the HR derived by a power-law spectrum with a photon index of 1.4-2. Some sources that exhibit extremely soft or hard features could be attributed to background galaxies with strong star formation or highly obscured background AGNs, which is common in X-ray surveys \citep[e.g.][]{2007ApJ...659...29K}.

\section{Searching for the Intracluster X-ray Population} \label{sec:ICX}

\begin{figure*}
    \epsscale{1}
    \includegraphics[width = 0.5\textwidth]{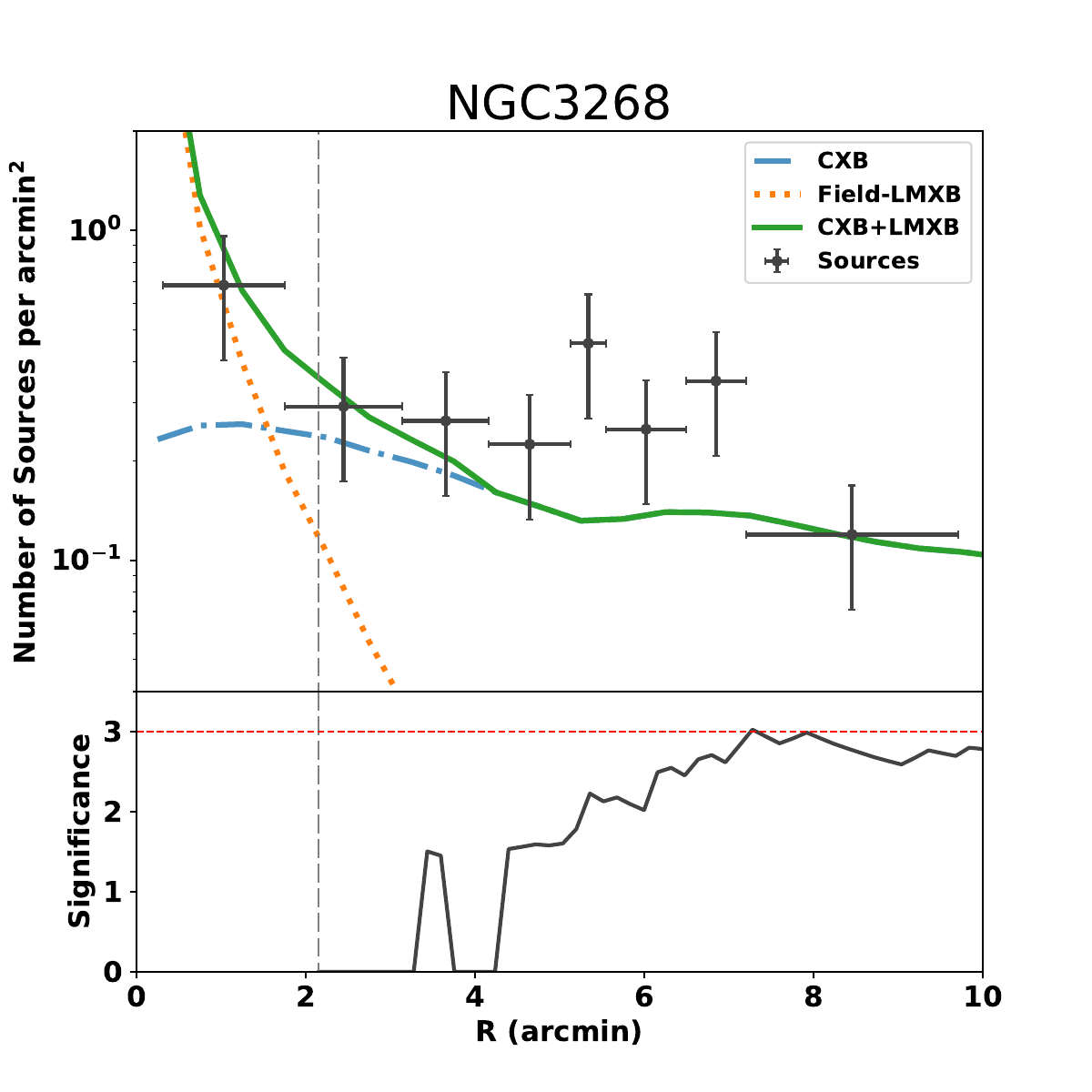}
    \includegraphics[width = 0.5\textwidth]{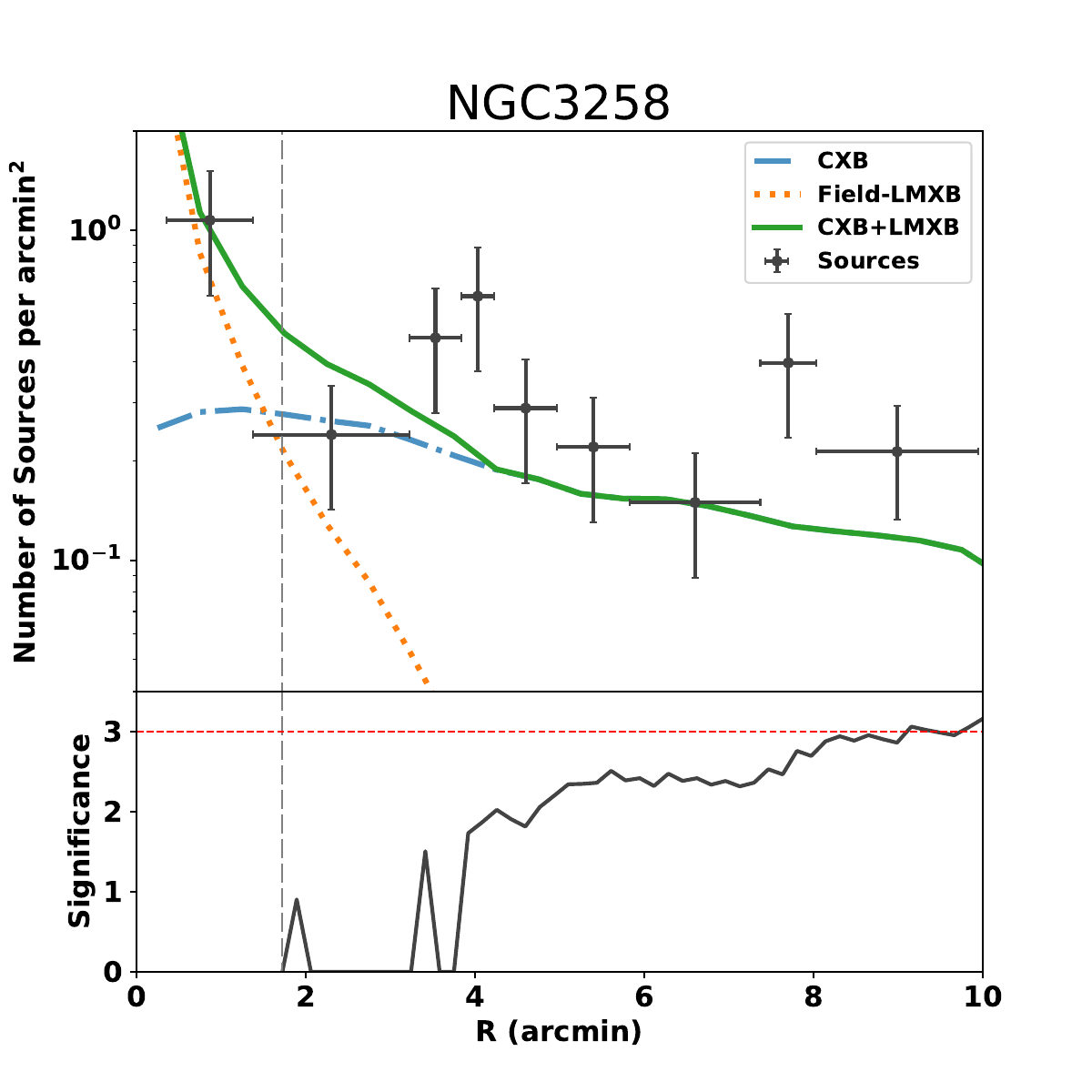}
    \includegraphics[width = 0.5\textwidth]{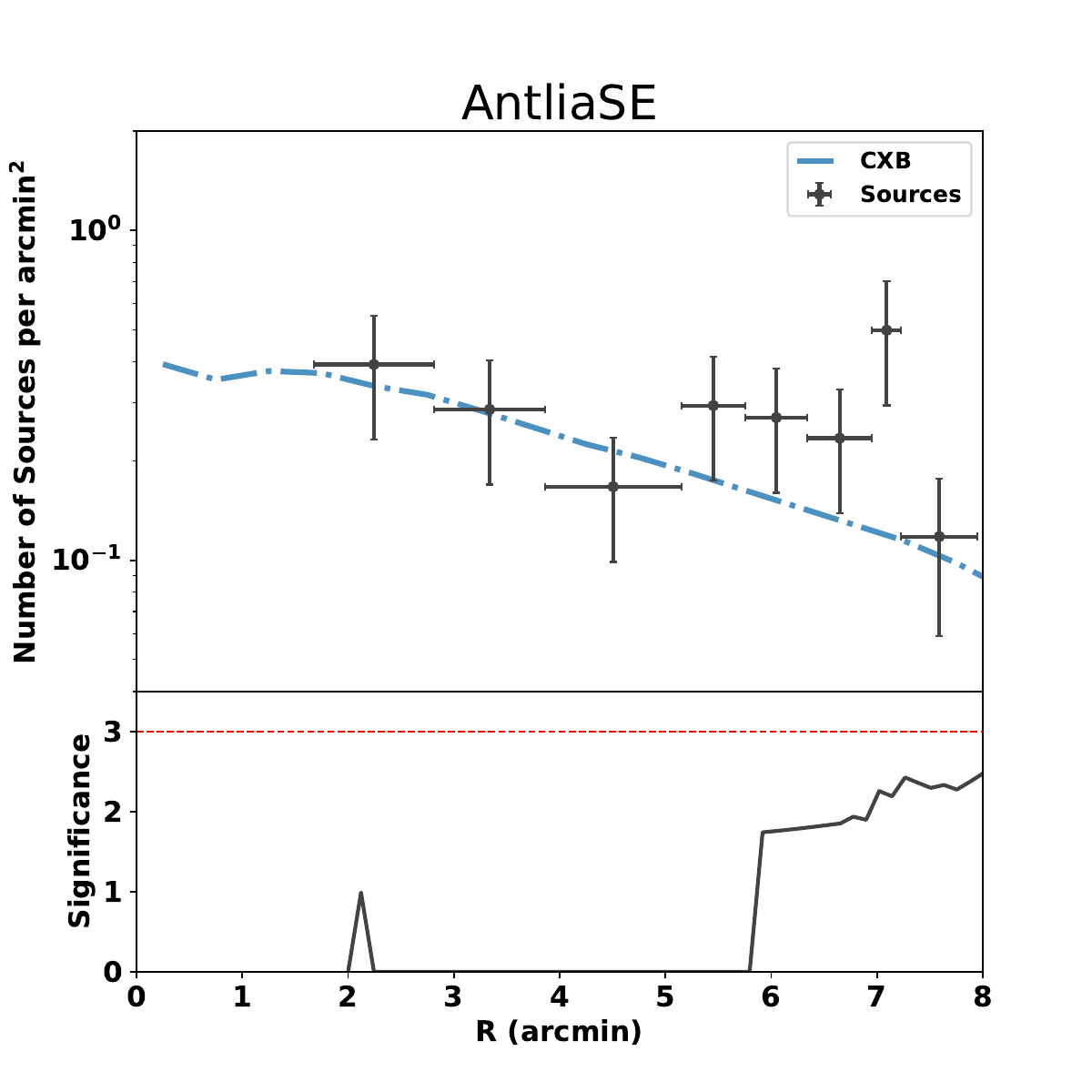}
    \includegraphics[width = 0.5\textwidth]{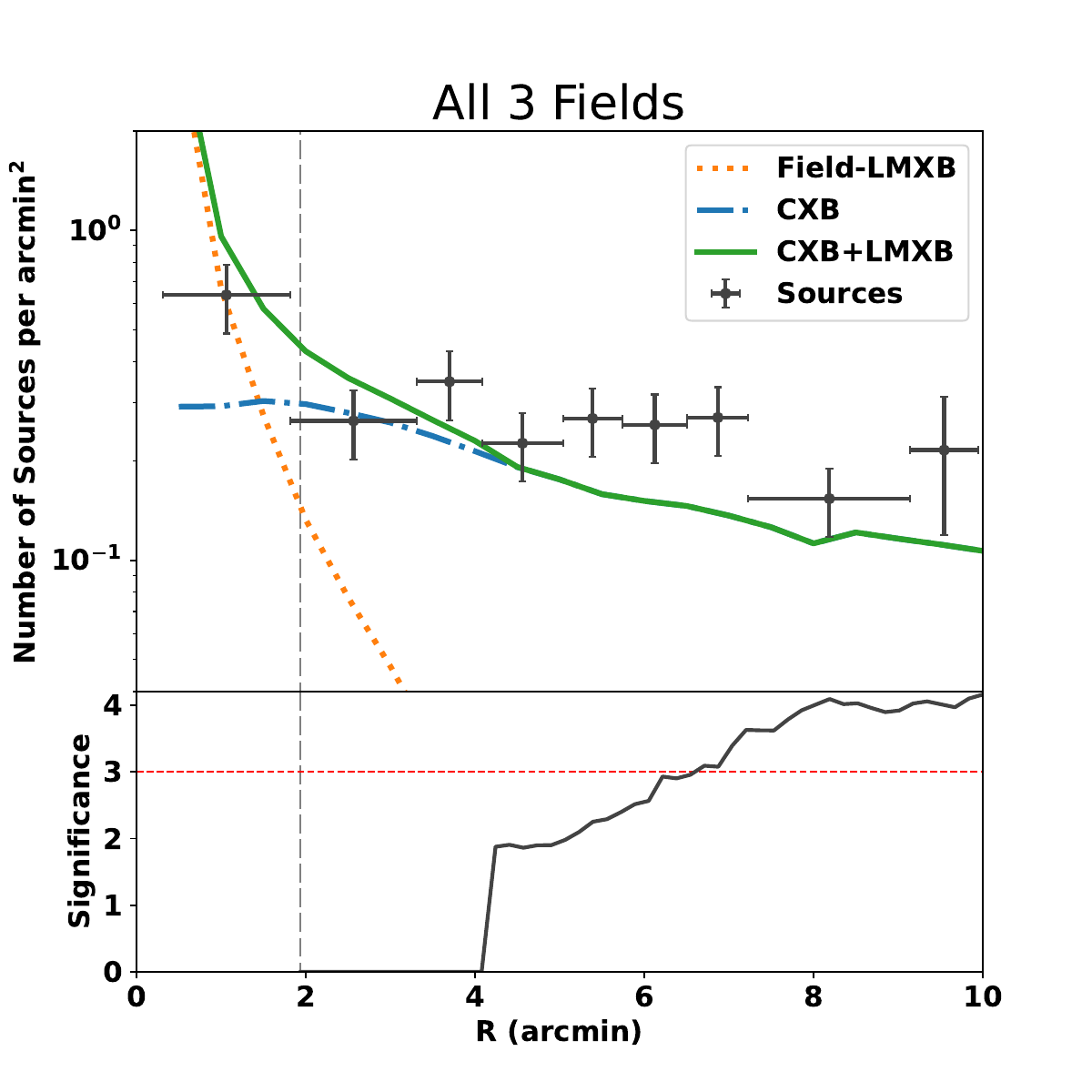}
    \caption{\textit{Upper left panel}: Surface number density distribution of the $F$-band  X-ray point sources in the NGC\,3268 field, excluding foreground stars, AGNs, and sources closer than 3 times the effective radius to satellite member galaxies. These point sources are adaptively binned to a minimum of 6 sources per bin (except for the last bin). The zero-point of the horizontal axis represents the center of NGC\,3268. The orange dotted line, blue dash-dotted line represent the prediction for field-LMXB and CXB components, respectively, while the green solid line shows their sum. A gray dashed line at $R=2.2'$ indicates three times the effective radius of NGC\,3268. The subplot below shows the significance of the excess of X-ray point sources calculated from beyond three times the effective radius. \textit{Upper right panel}: The legend is identical to the upper left panel, but displays the NGC\,3258 field. A gray dashed line at  $R= 1.7'$ marks the three times effective radius of NGC\,3258. \textit{Lower left panel}:  Distribution of sources surface density in the southeast field, centered on the aimpoint of the field-of-view. This field does not contain a BCG. Thus, the significance of the excess of  X-ray point sources is calculated  from the zero-point. \textit{Lower right panel}: Radial distribution of X-ray point sources after stacking the three fields, with each bin containing a minimum of 18 sources. The gray dashed line at $R=1.95'$ signifies three times the average effective radius of the two BCGs. The significance of the excess of X-ray point sources reaches a maximum of $4.2\sigma$.} 
    \label{fig:BCG_Xraysrc_Distribution}
\end{figure*}

ICX is defined as a kind of source that is located inside a cluster but unbound to the stellar component of member galaxies, which can be probed by the spatial distribution of X-ray sources \citep{2017ApJ...846..126H,2019ApJ...876...53J}. Figure \ref{fig:BCG_Xraysrc_Distribution} depicts the average surface density of the  $F$-band sources in different fields as a function of the projected radius $R$, excluding all foreground sources, AGNs and sources that are closer than three times effective radius  (taken from \citealp{2020MNRAS.497.1791C}) of satellite galaxies, which are marked with brown crosses in Figure \ref{fig:X-ray_map}.

There are two basic components: (i) field-LMXBs that trace the bulk stellar component, and (ii) the cosmic X-ray background (CXB). The radial distribution of the field-LMXBs is expected to tightly follow the bulk stellar mass distribution of the host galaxies, and the mass distribution follows that of the stellar light. The field-LMXB component is derived based on the stellar light distribution profile from \citet{2003A&A...408..929D}, after correcting the local X-ray detection limit. The stellar mass of the two BCGs are calculated according to \WISE \citep{2010AJ....140.1868W} $W1$ and $W2$ images, 
following the stellar mass-luminosity described in \citet{2019ApJS..245...25J}, and resulting in  $3.8\times10^{11}\ \Msun$ for NGC\,3268 and $2.1\times10^{11}\ \Msun$ for NGC\,3258.  The field-LMXB profile is built based on a group of annuli with a width of $0.5'$ centered at the BCG center. We derived the stellar mass and mean detection limit in each annulus and then estimate the number of field-LMXBs based on the X-ray source luminosity function \citep{2012A&A...546A..36Z}. 
Moreover, the CXB profile is derived through the following steps: First, we  calculate the median sensitivities of annuli with a width of $0.5'$  centered on the BCGs or the Antlia southeast field aimpoint. Then we derive the CXB surface density for each annulus using the $\log N$--$\log S$ relation  \citep{2007ApJ...659...29K}.  Finally, we combine all CXB density data points to obtain the final profile.

The number of excess sources is defined as $N_\mathrm{excess} = N_\mathrm{obs} - N_\mathrm{LMXB}-N_\mathrm{CXB} $, where $N_\mathrm{X}$ denotes the cumulative number of sources of the corresponding component. The statistical significance, as shown in the lower panel of Figure \ref{fig:BCG_Xraysrc_Distribution}, is calculated with $S = \sqrt{2}\ \mathrm{erf}^{-1} (1-Q(N,b))$, where $\mathrm{erf}^{-1}$ is the inverse of the error function, $Q(N,b) = e^{-b}b^N / N!$ is the Poisson probability of detecting $N$ sources given an expected background of $b$ sources, which refers to the sum of  field-LMXB and CXB in this case.  In order to remove the contamination of the BCGs, we only include the sources outside three times the effective radii of the two BCGs. However, the effective radii of the BCGs in previous studies (i.e. \citealp{2003A&A...408..929D} and \citealp{2020MNRAS.497.1791C}) are inconsistent, even if both are based on data from the Cerro Tololo Inter-American Observatory. This inconsistency can be attributed to the different definitions of galaxy edges caused by different exposures. Therefore, we measure the effective radii of the two BCGs with \WISE $W1$ images \citep{2010AJ....140.1868W}. The edge of a galaxy is defined as the $22\ \mathrm{mag \ arcsec^{-2}}$ (Vega magnitude) isophote in the $W1$ band, which corresponds to the 2.5$\sigma$ sky background level \citep{2019ApJS..245...25J}.  The effective radius of NGC\,3268 and NGC\,3258 is $0.72'$ ($7.2\ \kpc$) and $0.58'$ ($5.8\ \kpc$), respectively, and their three times the effective radii are depicted with gray vertical lines in Figure \ref{fig:BCG_Xraysrc_Distribution}.

\begin{figure}[ht!]
\epsscale{1.2}
\plotone{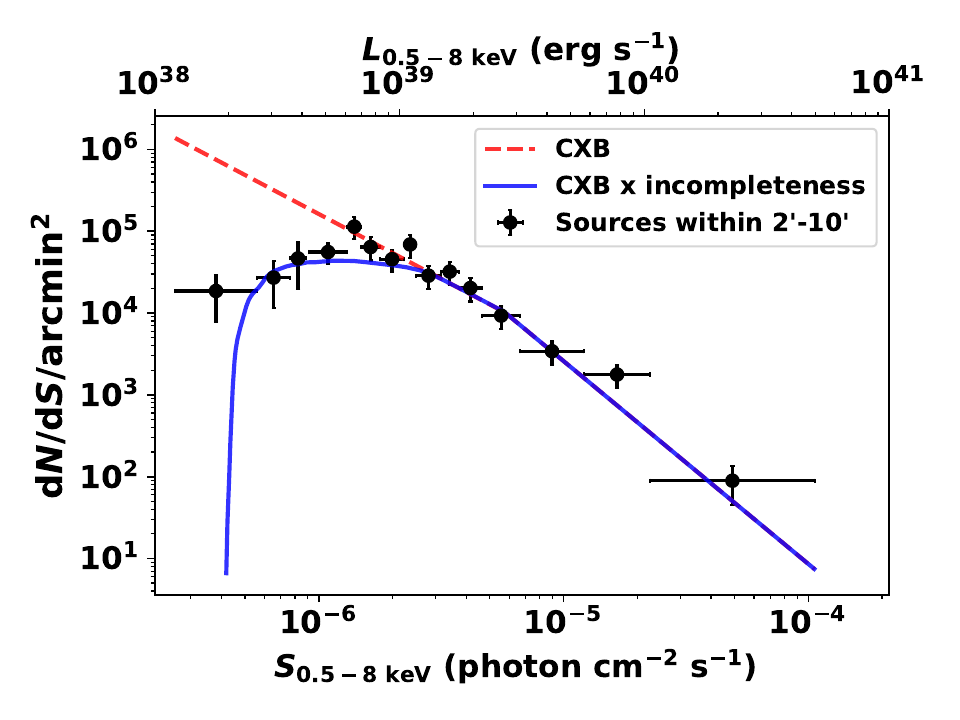}
\caption{The  logN–logS relation of the sources in the range of $2'<R<10'$, where the ICX is detected. The first three bins each has 3 sources, while the others each has 10 sources except  the last one.   The red dashed line is the empirical logN–logS model of CXB sources, and the blue solid line is corrected for incompleteness.  } \label{fig:logNlogS}
\end{figure}

The information regarding the excess is provided in Table \ref{table:ExcessInfo}. In the southeast field, the distribution begins at the aimpoint. However, in the NGC\,3268 and NGC\,3258 fields, the distributions  originate from the center of the BCG, which is why these two distributions extend up to $10'$ in radius. 
As a result, we find a statistically significant excess with $37.6$ excess sources, with a significance of $4.2\sigma$ in all three fields within a range of $3\bar{R_e}<R<10'$. Here $3\bar{R_e}=1.95'$ (20 kpc) indicates the three times the mean effective radius of the two BCGs, and $10'$ corresponds to 102 kpc. In Figure \ref{fig:logNlogS}, we plot the logN-logS relation of the sources within the range of $3\bar{R_e}<R<10'$, together with the CXB model \citep{2007ApJ...659...29K} corrected for detection incompleteness. The X-ray source surface density is higher than the CXB $\times$ incompleteness profile at $\sim2\times10^{-6}\ \mathrm{erg \ s^{-1}\ cm^{-2}}$. 
On the other hand, focusing on each field, we find  that NGC\,3258 field has 16.3 excess sources with a significance of $3.2\sigma$, and both the NGC\,3268 and NGC\,3258 fields have more excess sources compared to the southeast field that does not contain a BCG. 
Furthermore, combining these two fields, we obtain a statistically significant excess with $N_{\rm excess}=27.5$ and a significance level of $3.8\sigma$. This indicates a likely correlation between these excess X-ray point sources and the BCGs.

Additionally, we also consider the cosmic variance. Cosmic variance, $\sigma_c$, can be estimated using the following formula \citep[e.g.][]{1992ApJ...396..430L}:
\begin{equation}
    \sigma_c^2 = \frac{1}{\Omega^2}\int w(\theta)\mathrm{d}\Omega_1\Omega_2 = C_\gamma \theta_0^{\gamma-1}\Theta^{1-\gamma},
\end{equation}
here $w(\theta) = (\theta/\theta_0)^{1-\gamma}$ is the power-law angular two-point correlation function  \citep{1980lssu.book.....P}, $\theta_0\approx 0^\circ.00214$ is the correlation length according to a large survey based on \XMM data \citep{2009A&A...500..749E}, the constant $C_\gamma\approx2.25$  for the canonical value of  $\gamma=1.8$  \citep{1980lssu.book.....P}, $ \Theta=\sqrt{\Omega}=0.67\  \mathrm{deg}$ is the size of the field-of-view (FoV). The resulting cosmic variance is $\sigma_c = 0.15$. If we assume that the number of total CXB sources increases by $1\sigma_c$, resulting in 98.8 CXB sources, whereas the number of LMXB sources remains at 3.5, the remaining sources is $127 - 98.8 - 3.5 = 24.7$. As a result, the significance of the remaining sources is $3.1\sigma$, indicating that the excess still persists. For comparison, if the cosmic variance rises to $1.5\sigma_c$ and $2\sigma_c$, the statistical significance of the excess would drop to 2.6$\sigma$ and 2.3$\sigma$, respectively.
In conclusion, cosmic variance  is unlikely to be the primary origin of the excess sources.

\begin{deluxetable}{ccccccc}
\tablenum{2}
\tablecaption{Excess X-ray Sources in Different Fields}
\tablewidth{0pt}
\tabletypesize{\footnotesize}
\tablehead{
\colhead{Field}&    \colhead{Range} &  \colhead{$N_{\rm obs}$} &  \colhead{$N_{\rm LMXB}$} &   \colhead{$N_{\rm CXB}$} &  \colhead{$N_{\rm excess}$} &  \colhead{Significance} 
}
\decimalcolnumbers
\startdata 
3268     &   $3R_e\atob10'$ &  38 &    1.1 &  24.7 &  12.2 &  $2.8\sigma$ \\
3258     &       $3R_e\atob10'$ &  48 &  2.4 &  29.3 &  16.3 &  $3.2\sigma$ \\
3268 $+$ 3258 & $3\bar{R_e}\atob10'$     &   85 & 3.5 &  54.0 & 27.5 &   $3.8\sigma$\\
  SE    &   $0'\atob8'$   &    46 &  0 &   36.7 &  9.3 &   $2.3\sigma$ \\
  Total &  $3\bar{R_e}\atob10'$     &   127 & 3.5 &  85.9 & 37.6 &   $4.2\sigma$
\enddata 
  \tablecomments{ (1) Names of the fields. (2) Radial ranges of the excesses. The notation $3R_e$ indicates three times the effective radius. For NGC\,3268 and NGC\,3258, $3R_e$ is $2.2'$ (22 kpc) and $1.7'$ (17 kpc), respectively. The mean value, $3\bar{R_e}$, is $1.95'$ (20 kpc). The upper range extends to  $10'$ (102 kpc). (3) Number of observed X-ray sources. (4) Predicted number of field-LMXBs. (5) Predicted number of CXB sources. (6) Number of excess sources. (7) Significance of the excesses.  \label{table:ExcessInfo}}
\end{deluxetable}

\section{Discussion}\label{sec:discussion}

We detected a total of 202 X-ray point sources across three fields in the central $\sim200$ kpc region of the Antlia cluster. Beyond three times the mean effective radius of the two BCGs, we find a statistically significant excess at $4.2\sigma$ significance level with 37.6 sources, suggesting that these sources could be genuine ICX.  ICX has been identified in Virgo \citep{2017ApJ...846..126H} and Fornax \citep{2019ApJ...876...53J}, implying that the presence of these sources could be universal in nearby clusters.
By definition, ICXs are gravitationally unbound to any member galaxies, even though they could originate from a certain galaxy, akin to the so-called intracluster light (ICL) which is conventionally detected in optical/near-infrared band \citep{2022NatAs...6..308M}. In this section, we compare the ICX sources in Antlia with those in Virgo and Fornax, and discuss their potential origins.

\subsection{A Comparison with other clusters}

\citet{2017ApJ...846..126H}  have detected 116 excess sources  beyond $\gtrsim3R_e$ with a significance of $3.5\sigma$ using \Chandra observation of 80 Virgo ETGs. No significant excess has been found by these authors in their field analogs,  suggesting that the excess could have a cluster-related origin. Here we try to estimate the number of excess sources in Antlia based on Virgo data. Both surveys remove all sources within $3R_e$ from the member galaxies, which is why we just scale the areas of the footprints rather than the stellar masses. 
The detection limits of the Antlia and Virgo samples are different because of the distinction in instruments. The Virgo sample utilizes ACIS-S CCDs and include sources within $4'$ from the aimpoints, whereas the Antlia sample uses ACIS-I CCDs and accounts for sources within $8'$ from the aimpoints.
In order to correct the detection limit, we divide the Antlia sample into two annuli with ranges of $3\barRe<R<5'$ and $5'\leqslant R< 10'$, where $5'$  corresponds to a detection limit of $\sim 6\times10^{38}\ \ergs$, which is the second break of the field-LMXB X-ray luminosity function
\citep{2012A&A...546A..36Z}.  The areas of the inner and outer annuli are $2\times10^4\ \kpc^2$ and $4\times10^4\ \kpc^2$, respectively. The detection limit of the inner annulus is similar to the Virgo sample, while the outer one is expected to have $\sim 1/6$ sources according to the luminosity function. As a result, it is estimated that we can detect 25.7 and 7.3 excess sources in the inner and outer annuli. Thus, 33 excess sources are assumed to be detected based on Virgo data, which is similar to the real detection in Antlia. Note that all the galaxies in the Virgo sample are satellite galaxies, which means there is little stellar halo in the FoV. On the other hand, two BCGs are included in the Antlia sample. Thus,  Antlia BCGs may not have much stellar halo component. 

\citet{2019ApJ...876...53J} have constructed the source surface density profile out to a projected distance of $\sim30'$ ($\sim 180\ \kpc$)  towards the center of NGC\,1399, the BCG of the Fornax cluster. 
They have found an excess in the range of $7.5'\lesssim r\lesssim12.5'$ ($43.7\ \kpc \lesssim R\lesssim72.7\ \kpc$), with a  significance of $3.6\sigma$ and 183 excess sources. Interestingly, 109 excess sources have been believed to be related to the stellar halo around NGC\,1399, which was first discovered by \citet{2016ApJ...820...42I} based on the VLT Survey Telescope (VST). 
Here we also estimate the expected number of excess sources to be detected in Antlia with the Fornax result.  First, we calculate the masses of stellar halo in the Antlia and Fornax samples according to the radial surface brightness profile of NGC\,1399 stellar halo:  $\mu(R)=\mu_0+1.086(R/R_h)$, where  $\mu_0 = 23.4\ \mathrm{mag\ arcsec^{-2}}$ and $R_h=28\ \kpc$, and the total mass of the stellar halo is $4\times10^{11}\ \Msun$  \citep{2016ApJ...820...42I}.  This results in  stellar halo masses of $3\times10^{11}\ \Msun$ and $1.1\times10^{11}\ \Msun$  in the Antlia and Fornax samples, respectively.
Secondly, we scale the detection limit.
Due to the extensive coverage of the Fornax sample, we provide an overall estimation of Antlia FoV rather than dividing it into two parts, as mentioned earlier.
The median detection limit for the Fornax sample is $\sim 7.5\times10^{38}\ \mathrm{erg\ s^{-1}}$, which can lead to a detection of $\sim 9.4$ times more LMXBs according to the luminosity function of field-LMXBs. After scaling the mass of the BCG stellar halo and the detection limit,  it gives an expected number of 52.5 excess sources for the Antlia sample, which is more than our detection. 
This could also be attributed to the less stellar halo component around Antlia BCGs. 

In general, the discovery of the ICX in Antlia, Virgo and Fornax imply that these sources could be universal in nearby clusters. We find that the number of excess sources is similar to the sample of Virgo satellite galaxies, but it is fewer than the Fornax sample, which includes the BCG NGC\,1399. This could be due to the less stellar halo component around Antlia BCGs.

\subsection{The Origins of the Intracluster X-ray Population}

\subsubsection{X-ray sources associated with the intracluster light}
Intracluster light (ICL) is the stellar component which is stripped and expelled from member galaxies due to the their complex dynamic mechanisms, like frequent tidal interactions and mergers \citep{2021Galax...9...60C}. ICL 
can naturally form LMXBs because it is mostly composed of old stellar component, as indicated by its red color \citep{2017ApJ...834...16M}. Due to the lack of a clear boundary between the stellar halo and the ICL within galaxy clusters, \citet{2022NatAs...6..308M,2007ApJ...666..147G} have suggested that  stellar halo should be included in  ICL. Therefore, it is appropriate to include LMXBs associated with the BCG stellar halo as a subset of  ICL-LMXBs. Although current optical observations have not reported an additional component describing the stellar halo in Antlia, this could be due to the limited observation depth or the selection of models \citep[e.g.][]{2010ApJ...715..972J}. 
However, ICL-LMXBs are not likely to  dominate  the excess sources in the Antlia sample. A direct evidence is that the southeast field, that 
 does not contain a BCG, has  9.3 excess sources, which account for $76\%$ and $57\%$ of the excess sources in the NGC\,3258 and NGC\,3268 fields, respectively.  In general, although ICL-LMXBs could have a contribution, they are unlikely to dominate the ICX.

\subsubsection{GC-LMXBs}

Globular clusters (GCs) are more likely to harbor LMXBs (GC-LMXBs) because it is easier for compact objects to capture stars in a dense environment. 
However, GC-LMXBs do not have a large contribution to the excess of sources. 
We also estimate the number of GC-LMXBs in two annuli with ranges of $3\barRe<R<5'$ and $5'\leqslant R< 10'$, due to the break at $6\times10^{38}\ \ergs$ of the field-LMXB luminosity function \citep{2012A&A...546A..36Z}.  
According to the GC surface density profile in \citet[Figure 7]{2017MNRAS.470.3227C}, we estimate that $1330$ and $1660$ GCs falls in the inner annuli of the NGC\,3268 and NGC\,3258 fields, and 930 and 980 GCs are in the outer annuli of these two fields, respectively. Previous research shows that \citep{2001ApJ...557L..35A,2002ApJ...574L...5K,2006ApJ...647..276K,2006ARA&A..44..323F}
$4\%$--$5\%$ GCs harbor an LMXB down to a detection limit of $\logLx\geqslant37$ in nearby elliptical galaxies. The difference in detection limits leads to a detection of $\sim7\%$ and $\sim1\%$ GC-LMXBs in the inner and outer annuli.  Hence, there are approximately $4.9$ GC-LMXBs for the NGC\,3268 field (4.4 for the inner annulus and 0.5 for the outer one) and $6.0$ GC-LMXBs for the NGC\,3258 field (5.5 for the inner annulus and 0.5 for the outer one). In total, there is an estimated count of 10.9 GC-LMXBs, accounting for $\sim30\%$ of the detected excess sources. This estimation is consistent with the finding of \citet{2017ApJ...846..126H}, that $\sim30\%$ X-ray sources coincide spatially with GCs in the Virgo cluster. In a word, GC-LMXBs are unlikely to dominate the ICX.

\subsubsection{SN-kicked LMXBs}
An SN-kicked XRB is an XRB  that the compact object's progenitor experiences an anisotropic supernova explosion, imparting an initial kick velocity to the system \citep{1995MNRAS.274..461B} that could take it away from its birthplace. 
SN-kicked XRBs are most likely to be LMXBs \citep{1995MNRAS.274..461B,2008MNRAS.387..121Z}, as their lifetimes are longer than the timescale of migrating to the stellar halo. According to \citet{2013A&A...556A...9Z}, the number of X-ray point sources  beyond $4\atob10$ times the effective radius in ETGs of GC-LMXBs is approximate to that of  SN-kicked LMXBs. Therefore, in the Antlia sample, SN-kicked LMXBs may also have a similar contribution to GC-LMXBs.

\section{Summary} \label{sec:summary}

We  detect 202 sources in the inner $\sim$ 200 kpc region of the Antlia cluster based on \Chandra  observations, down to a $0.5$--$8\ \keV$ source detection limit of
$4.2\times10^{-7}\ \phcms$ ($2\times10^{38}\ \ergs$).  We present a source catalog and list their coordinates, multi-band fluxes and hardness ratios.
Also, we identify the optical and radio counterparts with \Gaia and \MeerKAT data. After that, we systematically analyze the $F$-band  X-ray source surface density distribution. The main results and conclusions are summarized as following:

i) We find a statistically significant  X-ray sources excess at a $4.2\sigma$ significance level with 37.6 excess sources in the range of $3\barRe<R<10'$, after stacking the X-ray source profiles of the three fields. These excess sources are not associated with the bulk stellar component; hence, they could be ICX unbound to any member galaxies.

ii) We find 12.2 and 16.3 excess sources in the NGC\,3268 and NGC\,3258 field within $3R_e<R<10'$, and 9.3 excess sources in the southeast field within $0<R<8'$. The fields containing a BCG have more excess sources. In addition, the significance of the excess sources in these two fields reaches $3.8\sigma$. These findings suggest a potential relation between BCGs and the ICX, perhaps due to the contribution of the stellar halo surrounding BCGs.

iii) Previous studies  found ICX in Virgo \citep{2017ApJ...846..126H} and Fornax \citep{2019ApJ...876...53J}, indicating that these sources could be universal in nearby clusters. After correcting the differences in footprints and detection limits, the number of excess sources in the Antlia sample is similar to that of the Virgo sample, but it is only one-third of the Fornax sample. This implies that Antlia BCGs may have less stellar halo component than Fornax BCG. 

iv) The origins of the excess sources could be ICL-LMXBs, GC-LMXBs and SN-kicked LMXBs. We suggest the LMXBs relating with the stellar halo surrounding BCGs should be included in ICL-LMXBs. GC-LMXBs and SN-kicked LMXBs, each component accounts for  $\sim30\%$ of the total excess sources.

\begin{acknowledgments}

Z.H. and Z.L. acknowledge support by the National Natural Science Foundation of China (grant 12225302) and the National Key Research and Development Program of China (NO.2022YFF0503402). 

Y.S. acknowledges support from Chandra X-ray Observatory grants GO1-22104X, GO2-23120X, and NASA grants 80NSSC22K0856.

M.H. is supported by the National Natural Science Foundation of China (12203001) and the fellowship of China National Postdoctoral Program for Innovation Talents (grant BX2021016). 

This paper employs a list of Chandra datasets, obtained by the Chandra X-ray Observatory, contained in~\dataset[DOI: 10.25574/cdc.222]{https://doi.org/10.25574/cdc.222}.

The \MeerKAT telescope is operated by the South African Radio Astronomy Observatory, which is a facility of the National Research Foundation, an agency of the Department of Science and Innovation.

This research made use of \texttt{photutils}, an \texttt{astropy} package for
detection and photometry of astronomical sources \citep{larry_bradley_2021_5796924}.

\end{acknowledgments}

 \bibliography{AN_IntraclusterSource.bib}{}
 \bibliographystyle{aasjournal}

\end{document}